\documentclass[preprint,12pt]{elsarticle}
\usepackage{amsmath, mathrsfs, latexsym, amssymb, color, graphicx, comment, hyperref, scrhack}
\usepackage{lmodern}
\usepackage{physics}
\usepackage{graphicx}
\usepackage{color}
\usepackage[usenames,dvipsnames]{xcolor}

\newcommand{\dfr}[2]{\frac {\displaystyle #1}{\displaystyle #2}}

\begin{document}

\begin{frontmatter}

\title{Berezinskii--Kosterlitz--Thouless transition close to zero temperature}

\author[label]{M.\,G.\,Vasin}
\address[label]{Vereshchagin Institute of High Pressure Physics, Russian Academy of Sciences, 108840 Moscow, Russia}

\begin{abstract}
The Berezinskii--Kosterlitz--Thouless (BKT) phase transition is considered in the condition of lowest temperatures, when thermal
fluctuations give place to quantum ones. For this goal, the critical dynamic of the Sine--Gordon model near the quantum critical point is considered. The approach based on the Keldysh--Schwinger technique of non-equilibrium dynamics description is used, as well as the method of taking into account the crossover from the thermal fluctuation regime to the quantum one in the renormalization group technique. For the system with low vortex concentration, it is shown that the BKT transition unavoidably occurs at a temperature above the crossover temperature from the thermal fluctuation regime to quantum one. As a result at small temperatures the critical exponent of the BKT transition, $\eta=1/4$, remains unchanged, however, the universal relation of the superfluid density jump adjusts a little from the well-known relation by Nelson and Kosterlitz~\cite{NK}.
\end{abstract}


\end{frontmatter}

\section{Introduction}

It is well known that in two-dimensional systems with continuous symmetry of the order parameter at finite temperatures, there can be no long-range translational order. This is due to the destruction of the long-range order by topological vortices resulting from thermal fluctuations of the system. However, such systems can undergo the topological Berezinskii--Kosterlitz--Thouless phase transition which occurs in the vortices subsystem. Berezinskii--Kosterlitz--Thouless (BKT) works\,\cite{Berezinskii,KostThoul,KT1972} brought a paradigm of topological phase transitions driven by topological excitations to prominence in the condensed matter physics. The binding--unbinding BKT transition separates the low-temperature, $T<T_{BKT}$, the confined phase where topological excitations (vortices) of the opposite sign are bound into `neutral' dipoles, and the high-temperature unbound phase, at $T>T_{BKT}$, where topological excitations unbind loose and form a `free' neutral plasma.

By now, the BKT transition was found in several 2D systems. For example, in the $^4$He films \cite{Bishop}, in the Josephson junction arrays, thin disordered superconducting granular films \cite{Fazio,VB}, or in a trapped quantum degenerate gases \cite{Z}. In some of them, like the 2D superconductor-insulator transition, the BKT transitions metal-insulator and metal-superconductor take place at very small temperatures, which allows supposing these BKT transitions cross in a quantum critical point at $T=0$ \cite{VB,Shon}. In \cite{Shon} it was shown that at $T=0$ the universality class of the BKT transitions does not change. However, the question of how the continuous crossover from the thermal to the quantum ones influences the BKT transition in this case.

In the case of second-order phase transitions, it was experimentally found that at ultra-small phase transition temperatures the quantum-to-classical crossover (QCC) takes place, which appears in the change of critical exponents near the quantum critical point. In the 3D systems, the critical exponents continuously approach the mean-field ones at $T\to 0$ \cite{Steijger,Erkelens,Stishov2,Stishov1,Bittar,VVR,PhysA21}. The reason for this change is that the values of the critical exponents depend on the nature of the critical fluctuations. Accordingly, the change in the critical exponents reflects the crossover from the strong thermal fluctuations to the weak quantum fluctuations mode, the system's universality class remaining intact.

However, in the case of the BKT transitions in 2D systems, this quantum-to-classical crossover was not explored. One can suppose at zero temperature a quantum phase transition can occur in such systems, in which the resulting long-range order exists due to the absence of thermal and weakness of quantum fluctuations. In the quantum regime, the system effective dimension becomes over $2$, and one can expect that system critical behavior will suffer considerable changes.

This raises the investigation of the quantum-to-classical crossover in 2D systems is of special interest. In the present paper, this subject is discussed in terms of the critical dynamics approach, used in previous works for the description of QCC in 3D systems \cite{VVR,PhysA21}.

\section{Critical dynamics of Berezinskii--Kosterlitz--Thouless phase transition}

It is well known that quantum mechanics is dynamic in itself since it considers time as an additional equal in the rights dimension of the system, which naturally determines the critical exponents. Therefore, the key issue for solving our problem is a description of critical dynamics in classical and quantum cases as corresponding limits in terms of common dynamical theory. For this, we use the Keldysh--Schwinger technique \cite{Keldysh,Schwinger,Kamenev} in the same way as it was done in \cite{VVR,PhysA21} for a second-order phase transition.

\subsection{Model}

Let us consider a system described by the XY-model in equilibrium state.
The Hamiltonian has the following form:
\begin{gather*}
H=\sum\limits_{\bf r,l}\mathcal{E}S^2\left[1-\cos( \delta\Phi_{\bf r,\,l})\right],
\end{gather*}
where ${\bf r}$ is a lattice point, ${\bf l}$ are neighboring lattice points, $\mathcal{E}$ is the elastic (exchange) interaction energy.

The statistical mechanics of the considered system can be described in the continuous presence in terms of Sine--Gordon model (see Appendix I).
The model partition function can be written as follows:
\begin{gather*}
W=\int \mathfrak{D}A \exp \left[ -\dfr{\tau}{\hbar}H(A)\right],
\end{gather*}
where $A$ is an ancillary field, $\tau$ is the characteristic time, $\hbar$  is the reduced Planck constant, and the corresponding Hamiltonian of this model has the following form:
\begin{gather*}
H=\hbar\tau^{-1}\int\limits_V\mathrm{d}^2{\bf r}
\left[\dfr{\hbar\tau^{-1}}{4\mathcal{E}} A\nabla^2A-\dfr{g}{a^2}\cos(2\pi A)\right],
\end{gather*}
where $a$ is the characteristic space scale, $g=e^{-\mathcal{E}_c\tau/\hbar}$ is the vortex areal density, and $\mathcal{E}_c$ is the vortex core energy (see appendix I). Below we will consider only case of low-density vortex gas  $g\ll 1$ \cite{Minnhagen,Ryzhov}, in which the BKT transition takes place.

Below it will become clear that the definition of the characteristic time, $\tau$, plays a very important role near the zero-point. Therefore, we should devote this subject's attention.

According to the third law of thermodynamics, a system at absolute zero temperature exists in its ground state. At the same time from quantum mechanics, we know that a quantum system cannot have exactly zero energy. This energy gap is the result of the zero-point fluctuations of the system which frequency is the smallest fluctuation frequency of the system $\omega_0\neq 0$. It is naturally supposed that the fluctuation with the smallest frequency corresponds to the spin-wave which half-length is equal to the system linear size. It is approximately the energy of turn the spin on the angle $\pi$, therefore $\hbar\omega_0\approx \mathcal{E}$.

In overhand, according to the Heisenberg uncertainty principle, the minimal time that is necessary to determine energy with accuracy $\Delta E$ is $\tau\geqslant \hbar/\Delta E$. This value defines the time scale of the system. The energy uncertainty corresponding to the quantum fluctuations cannot be less than the ground state energy  $\Delta E\geqslant E_0=(\hbar\omega_0/2)\,\mbox{cth}\left(\hbar\omega_0/{2k_bT}\right)$, where $\omega_0$ is the minimal ground state frequency.
Therefore $\tau=2/\omega_0\,\mbox{cth}\left(\hbar\omega_0/{2k_bT}\right) $ (see appendix II).
From the last expression, one can see that at the high temperatures the characteristic time scale is $\tau\sim\hbar/k_bT$, however, at small temperatures, it does not diverge to infinity but seeks to the finite value $\tau\sim 2/\omega_0$ defined by the frequency of the zero-point fluctuations.

Above becomes principally important for understanding the critical behavior of the system close the zero-point since in the frequency region $\omega<\omega_0$ the fluctuations do not exist and undo calculation of infrared divergences in the renormalization constants one should consider the $\omega\to\omega_0$ limit instead of the usual one, $\omega\to 0$.

\subsection{Application of the Keldysh--Schwinger technique to Sine--Gordon model}

The description of non-equilibrium effective field theories needs the doubling of the degrees of freedom. The point is the case of a dissipative system its action depends on the initial state of the system and, accordingly, on the choice of the initial time. The averaging operation is not defined in this case and, as a consequence, the statistical theory can not be formulated. To get around this problem, one uses the following approach: name our field as $A^+$, and consider an additional copy of our system, with the same transition amplitude, and the field in the replica system $A^-$. Recall that both fields are, in fact, identical,  hence $\langle A^+(0) | A^-(0 )\rangle = 1 $. Using these two fields, we inverse time in the second system and close the integration contour at $ t = \infty $, where the system is in an equilibrium state which is characterized by some energy distribution function $f(\omega)$. If the system described by the $A$ field obeys Bose-Einstein statistics, then $f(\omega)=\mbox{cth}\left({\hbar\omega}/{2k_bT}\right)$. After that, we carry out the Keldysh rotation  \cite{Keldysh,Kamenev}, going to the new fields: $A^q=(A^+-A^-)/\sqrt{2}$; $A^{cl}=(A^++A^-)/\sqrt{2}$, and write the path integral of the dynamic Sine--Gordon theory as:
\begin{multline*}
\mathcal{W}=\int \mathfrak{D}A^{cl} \mathfrak{D}A^{q}\exp \left[-\int\limits_{0}^{\infty}\mathrm{d}t\int\limits_V\mathrm{d}^2{\bf r}
\,\left\{\bar A\hat G^{-1}\bar A+\right.\right.\\ 
\sigma\sin(\sqrt{2}\pi A^{cl})\sin(\sqrt{2}\pi A^q)\Big\}\Bigg],
\end{multline*}
where $\bar{A}=(A^{cl},\,A^q)$, $\int\mathfrak{D}A$ denotes a functional integration, $\sigma=e^{-\mathcal{E}_c\tau/\hbar}/a^2\tau$, and $\hat G^{-1}$ is the $A$-field inverse Green function operator.
Below, the path integral will be convenient to represent it in a pulse-frequency representation, in which the functions of sine in the action can be represented as a power series over the field $A$. Thus
\begin{multline*}
\mathcal{W}=\int \mathfrak{D}A^{cl} \mathfrak{D}A^{q}\exp \left[-\int\limits_{0}^{\infty}\dfr{\mathrm{d}\omega}{2\pi}\int\limits_{V^{-1}}\dfr{\mathrm{d}^2{\bf k}}{(2\pi)^2}
\,\left\{\bar A_k\hat G^{-1}\bar A_{-k}+
\sigma (\sqrt{2}\pi)^2A^{cl}_kA^q_{-k}
\right.\right.\\\left.
-\dfr{\sigma(\sqrt{2}\pi)^4}{3!}\iint\dfr{\mathrm{d}^3p_1}{(2\pi)^3}\dfr{\mathrm{d}^3p_2}{(2\pi)^3}
\left( A^{cl}_{k}A^{cl}_{p_1}A^{cl}_{p_2}A^q_{-k-p_1-p_2}
+A^{cl}_{k}A^{q}_{q}A^{q}_{p_2}A^q_{-k-p_1-p_2}\right)\right\}
\\ \left.
-\dfr{\sigma(\sqrt{2}\pi)^4}{(3!)^2}\int\limits_{-\infty}^{\infty}\mathrm{d}t\int\limits_V\mathrm{d}^2{\bf r}(A^{cl}A^q)^3e^{i{\bf kr}+i\omega t}+\ldots\right],
\end{multline*}
where the $A$-field Green function operator in reciprocal space has the following form:
\begin{equation}\label{GF}
\hat G=\left[ \begin{array}{cc} \displaystyle G^K & G^R \\[12pt]
\displaystyle G^A &  0 \end{array}\right]=\left[ \begin{array}{cc} \displaystyle \dfr{\alpha\omega \,\mbox{cth}\left({\hbar\omega}/{2k_bT}\right)\Pi^{\omega_M}_{\omega_0}(\omega) }{\lambda^2{\bf k}^4+\alpha^2\omega^2} & \displaystyle \dfr1{\lambda{\bf k}^2-i\alpha\omega} \\[12pt]
\displaystyle \dfr1{\lambda{\bf k}^2+i\alpha\omega} &  0 \end{array}\right],
\end{equation}
where $\lambda=\hbar/4\mathcal{E}\tau^{2}$, $\alpha=a^{-2}$, and $\Pi^{\omega_M}_{\omega_0}(\omega)=\theta(|\omega|-\omega_0)\theta(\omega_M-|\omega|)$ is the product of the Heaviside functions, where $\omega_0$ and $\omega_M$ are the zero fluctuations frequency and the Debye frequency respectively (see Appendix II).

\subsection{Critical dynamics close to zero temperature}

The critical behavior of the system can be considered within the critical dynamics technique \cite{Halperin,Vasiliev}. The critical dynamics rests on the hypothesis of dynamical scaling, according to which the action should be invariant under scale transformations, which conformally expand the space and time coordinates ($\omega \sim k^{d_{\omega}}$). In this case the summarized dimension, $d=d_k+d_{\omega}$ ($d_{\omega}=z$ is the dynamic exponent), has the same role as the conventional (momentum) dimension, $d_k$, in the static case. The canonical dimensions of the fields and model parameters are determined from the condition of dimensionless action. The corresponding summarized canonical dimensions, $d[F]$, of any values, $F$, are defined as
\begin{gather*}
d[F]=d_k[F]/z+d_{\omega}[F],
\end{gather*}
where $d_{\omega}[F]$ is the frequency dimension \cite{Vasiliev}. The canonical dimensions of the values of our theory are given in the following table:
\begin{equation*}
\begin{tabular}{|c|ccccccc|}
  \hline
  $F$&$k$&$\omega$&$\lambda$&$\alpha$&$A^{cl}$&$A^q$&$\sigma$\\ \hline
  $d_k[F]$&$1$&$0$&$0$&$2$&$-2$&$-2$&$2$     \\
  $d_{\omega}[F]$&$0$&$1$&$1$&$0$&$-1$&$-1$&$1$     \\
  $d[F]$&$1/2$&$1$&$1$&$1$&$-2$&$-2$&$2$   \\ \hline
\end{tabular}
\end{equation*}
The renormalization procedure only refines these values, which leads to the replacement of the canonical dimensions by the critical ones.
It is carried out with the standard method. It is assumed that the fields $A^q$, $A^{cl}$ are slow-varying ones, such that the Fourier-transformed fields have only long-wave components: $|{\bf k}|<|{\bf k}^*|$; $\omega < \omega^*$. In the first step of the RG transformations one integrates the partition function over the components of the fields in the limited wave band $|{\bf k}^*| <|{\bf k}|<|{\bf k}^*|\Delta^{1/z}$, $\omega^*<\omega<\omega^*\Delta$, where $\Delta \gtrsim 1$ is the regularization parameter (cutoff of the momentum), and ${\bf k}^*\to 0$, $\omega^*\to 0$. It is expected that under certain conditions this action has a structure similar to the original one, in this case a model is multiplicatively renormalizable. One can check that the formulated model satisfies this criterion. As a result one gets an effective action $S_{\Delta}$ with renormalized parameters ($Z_{\lambda}$, $Z_{\alpha}$, $Z_{\sigma}$), which are named ``constants of renormalization'' and depend on the cutoff $\Delta$. In the second step, one makes the inverse scaling transformation of the fields ($Z_{A^q}\sim\Delta^{-d[A^q]}$, $Z_{A^{cl}}\sim\Delta^{-d[A^{cl}]}$) and coordinates which is aiming to
restore the original cutoff scale ${\bf k}^*$ and $\omega^*$. Then the renormalized parameters have the following form:
\begin{equation*}
\begin{array}{l}
\lambda^{(R)}=Z_{\lambda}Z_{A^q}
Z_{A^{cl}}Z_k^2\Delta^{-1-2/z}=Z_{\lambda}\Delta,\\
\alpha^{(R)}=Z_{\alpha}Z_{A^q}Z_{A^{cl}}Z_{\omega}\Delta^{-1-2/z}=Z_{\alpha}\Delta,\\
\sigma^{(R)}=Z_{\sigma}Z_{A^q}Z_{A^{cl}}\Delta^{-1-2/z}=Z_{\sigma}\Delta^{2}.
\end{array}
\end{equation*}
\begin{figure}[h]
\centering
   \includegraphics[scale=0.2]{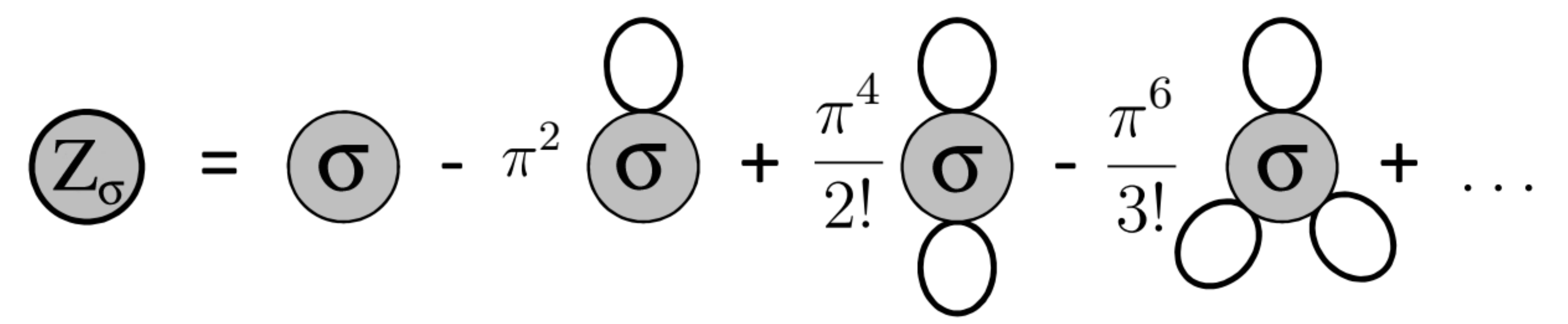}
   \caption{The first-order representation of the renormalization constant of the $\sigma $ parameter in the Feynman diagrams\,\cite{Min}. The loop corresponds to the integration of the Keldysh part of the correlation function, $G^K$, over $k=({\bf k},\,\omega )$ and is calculated as $\Omega (\Delta)$~(\ref{loop}). }
   \label{F1}
\end{figure}

Let us consider the case of the low-density vortex gas\,\cite{Minnhagen,Ryzhov}. The $\lambda$ renormalization is the same that in the static theory, and the $\alpha$ renormalization is trivial (see Appendix III).
The only nontrivial is the renormalization of $\sigma$.
In Fig.\,\ref{F1} in graphical form this renormalization is presented. After summarizing of the infinite series we have
\begin{gather*}
\sigma^{(R)}=Z_{\sigma}\Delta^2\approx\sigma\exp\left[-\pi^2 \Omega(\Delta)\right]\Delta^2,
\end{gather*}
where $\Omega(\Delta)$ is the divergent contribution of the $A^q$-field propagator loop:
\begin{multline*}
\Omega(\Delta)=\int\limits_{k^*}^{k^*\Delta}\dfr{\mathrm{d}^3k}{(2\pi)^3}\langle A^{cl}A^{cl}\rangle_{k}=\\
2\int\limits_{V^{-1}}\dfr{\mathrm{d}^2{\bf k}}{(2\pi)^2}\int\limits_{\omega^*}^{\omega^*\Delta}\dfr{\mathrm{d}\omega}{2\pi}\,\dfr{\alpha\omega \,\mbox{cth}\left({\hbar\omega}/{2k_bT}\right)\Pi^{\omega_M}_{\omega_0}(\omega)}{\lambda^{2}{\bf k}^4+\alpha^2\omega^2}=\\
\dfr{1}{4\lambda}\int\limits_{\omega^*}^{\omega^*\Delta}\dfr{\mathrm{d}\omega}{2\pi}\,\mbox{cth}\left({\hbar\omega}/{2k_bT}\right)\Pi^{\omega_M}_{\omega_0}(\omega).
\end{multline*}
Taking into account the conditions imposed by the $\Pi^{\omega_M}_{\omega_0}(\omega)$ function one should limit the lowest integration limit on frequency by the $\omega_0$ value, therefore
\begin{gather}\label{loop}
\Omega(\Delta)=\dfr{1}{4\lambda}\int\limits_{\omega_0}^{\omega_0\Delta}\dfr{\mathrm{d}\omega}{2\pi}\,\mbox{cth}\left({\hbar\omega}/{2k_bT}\right).
\end{gather}

In the high temperatures limit, $\hbar\omega\ll 2k_bT$, i.e. when $\mbox{cth}\left({\hbar\omega}/{2k_bT}\right)\propto 1/\omega$, this integral logarithmically diverges at small frequencies. This limit corresponds to the usual BKT transition. However, when $\hbar\omega\approx {2k_bT}$ the correction related to the thermal fluctuations attenuation appears. The integral divergence weakens, and at $T=0$ vanishes at all.

The calculation of like these integrals, which taking into account the crossover from thermal fluctuations regime to quantum fluctuations one, earlier was considered in \cite{VVR,PhysA21}. The details are presented in Appendix IV, and the result can be presented as follows:
\begin{gather*}
\Omega(\Delta)\approx \dfr{\omega_0}{8\pi\lambda}
\dfr{\mbox{cth}\left(x_0\right)}{\Lambda(x_0)}\left(e^{\Lambda(x_0)\ln\Delta}-1\right) ,
\end{gather*}
where
\begin{gather*}
\Lambda (x_0) = 1-2x_0\,\mbox{csch}(2x_0),
\end{gather*}
and $x_0=\omega_0{\hbar}/{2 k_bT}=\mathcal{E}/{2 k_bT}$.

Thus, the renormalization of $\sigma$ has the following form:
\begin{multline*}\label{RenG}
\sigma^{(R)}=\sigma\exp\left[-\pi^2\Omega(\Delta)\right]\Delta^2\\
\approx \sigma\exp\left[2\ln\Delta-\dfr{\pi\omega_0}{8\lambda}\dfr{\mbox{cth}(x_0)}{\Lambda(x_0)}
\left(e^{\Lambda(x_0)\ln\Delta}-1\right)\right].
\end{multline*}

 Taking into account that $\hbar\omega_0\approx \mathcal{E}$, $\lambda=\hbar/4\mathcal{E}\tau^{2}$, and $\tau=2\omega_0^{-1}\mbox{cth}^{-1}\left(\hbar\omega_0/{2k_bT}\right) $ one can see that at the relatively high temperatures, $2{k_bT}\gg \mathcal{E}$, the $\sigma$ renormalization has the following form:
\begin{gather*}
\dfr{d\ln \sigma}{d\ln\Delta}\approx  2-\dfr{\pi\omega_0}{8\lambda}\mbox{cth}\left(\mathcal{E}/2{k_bT}\right)
e^{\Lambda(x_0)\ln\Delta}\approx 2\left(1-\dfr{\pi}2\dfr{\mathcal{E}}{k_bT}\right).
\end{gather*}
This gives a well-known from school books value of BKT-transition temperature: $T_c=\pi\mathcal{E}/2 k_b$, at which the sine in the Hamiltonian becomes non-relevant. However, in the ultralow temperatures limit the phase transition temperature is a little adjusted: $T_c=\mathcal{E}/2k_b\,\mbox{arccth}(\pi)$.

Following usual whey one can come to the system of renormalization group equations is written as follows:
\begin{equation}\label{RenGroup}
\begin{array}{l}
\displaystyle\dfr{d\ln \sigma}{d\ln\Delta}\approx 2-\dfr{\pi\omega_0}{8\lambda}\mbox{cth}\left(\hbar\omega_0/{2k_bT}\right)
e^{\Lambda(x_0)\ln\Delta}
=2-\dfr{2\pi}{\mbox{cth}\left(\mathcal{E}/{2k_bT}\right)},\\[12pt]
\displaystyle\dfr{d\lambda}{d\ln\Delta}\approx \dfr{\sigma^2}{2\lambda }\mathcal{H}_0\{r^2\}_{k^*},\quad
\dfr{d\ln\alpha}{d\ln\Delta}\approx 1,
\end{array}
\end{equation}
where $\mathcal{H}_0\{r^2\}_{k^*}$ is the is Hankel transformation of the $r^2$ function in the $k^*$ point, diverging in the thermodynamic limit.
This divergence is excluded by the redefining $\sigma$ and $\lambda$ as $\sigma\to\sigma'=\sigma\sqrt{\mathcal{H}_0\{r^2\}_{k^*}}/\omega_0\,\mbox{cth}
\left(\mathcal{E}/{2k_bT}\right)$, $\lambda\to\lambda'=8\lambda /\pi\omega_0\,\mbox{cth}\left(\mathcal{E}/{2k_bT}\right)$. As a result (\ref{RenGroup}) takes the following form:
\begin{equation*}
\displaystyle\dfr{d\sigma'}{d\ln\Delta}\approx \left(2-\dfr{1}{\lambda'}\right)\sigma'
,\quad
\displaystyle\dfr{d\lambda'}{d\ln\Delta}\approx \sigma'^2,\quad
\dfr{d\ln\alpha}{d\ln\Delta}\approx 1,
\end{equation*}
where the exponential factor in the first expression is deemed as a unity, since $\Delta\approx 1$. The third expression is trivial and can be dropped out. Introducing more convenient variables for this purpose as $z={\sigma'}$ and $y=2-1/\lambda'$ (as a result $1/\lambda'=2$ at $T=T_c$) we come to the well-known form of BKT-model renormalization group:
\begin{gather*}
\displaystyle\dfr{dz}{d\ln\Delta}\approx 2yz
,\quad
\dfr{dy}{d\ln\Delta}\approx z^2.
\end{gather*}

The critical exponent $\eta $ corresponding to the correlation length, $\langle SS\rangle_{\bf r}\propto |{\bf r}|^{-\eta}$, is defined as, $\eta=\lambda'/2$~\cite{NK}. It means that
\begin{gather*}
\eta=\dfr{\lambda'}2=\dfr{4\lambda}{\pi\omega_0}\mbox{cth}^{-1}\left(\mathcal{E}/{2k_bT_c}\right)=\dfr1{4\pi}\mbox{cth}\left(\mathcal{E}/{2k_bT_c}\right)=\dfr14,
\end{gather*}
and the critical exponent does not depend on the proximity to the zero-point.

\subsection{Superfluid density jump}

Usually, in the phase transition description, we are interested in critical exponents, which are universal and measurable. Unfortunately, in the case of BKT-transition, such critical exponents are practically absent, except the critical exponent corresponding to the correlation length, $\eta $. From the above one can see that in BKT transition this exponent does not change at the ultrasmall temperatures, $T\to 0$. Therefore the quantum-to-classical crossover in the considered system cannot be observed in the same form as in the second-order phase transition \cite{VVR,PhysA21}. Moreover, according to the renormalization group equation system, the temperature dependence of the relaxation time also does not change at this crossover.

However, the quantum-to-classical crossover can be manifest in BKT-transition nevertheless. For this one should consider the superfluid density $\rho $ jump. This jump is universal and manifests the asymptotic value of the superfluid density drops to zero discontinuously at $T=T_c$, which means vortices are irrelevant for $T<T_c$.

It is known \cite{Berezinskii} that the parameters of the Bose gas be related to those of the magnetic system as:
\begin{gather*}
\mathcal{E}=\dfr{\rho \hbar^2}{m},
\end{gather*}
where $m$ is the effective mass. In the critical point the superfluid density changes from some value $\rho_c$, corresponding to the superfluid liquid, to zero, $\rho=0$, corresponding to the normal state. Using
\begin{gather*}
\rho =\dfr{\mathcal{E} m}{\hbar^2}=\mbox{arccth}(\pi)\dfr{2k_bT_cm}{\hbar^2}.
\end{gather*}
This is different from the classical case, where the factor before the fraction is $1/\pi$. Thus, we come to the relation $\lim\limits_{T\to T_c} \rho(T)/T=3.65\times10^{-9}$ ($g/cm^2\,K$), which adjusts a little the well-known relation $\lim\limits_{T\to T_c} \rho(T)/T=3.52\times10^{-9}$ ($g/cm^2\,K$) by Nelson and Kosterlitz~\cite{NK}.

The physical meaning of this behavior is quite obvious: when the temperature becomes so low that the system is subject only to zero quantum fluctuations, then the superfluid density is determined only by them, and for this reason, the superfluid density jump at the BKT-transition does not depend on temperature. At the same time, the universality of the transition is preserved, and the introduced in \cite{NK} critical exponent, $\eta=1/4$, remains unchanged.

It can be assumed that the above-found deviation of the temperature dependence of the superfluid density jump value for BKT-transition from the linear one can be observed in ultracold 2D XY-spin or Bose gas systems.

\section{Conclusions}

The short formulation of the result of the presented analysis is the statement of impossibility of the BKT transition at $T=0$ in the low-density vortex gas.  The BKT transition unavoidably occurs at the temperature $T_c\approx \pi\mathcal{E}/2k_b$ which is higher the crossover temperature from the thermal to quantum fluctuation regimes, $T_{0}\approx\mathcal{E}/2k_b$ \cite{PhysA21}. It explains the results of the above analysis, which shown the immutability of the critical behavior of the BCT transition at $T\to 0$. In the temperature interval $T_c>T>T_0$ the system can be only in the bonded vortices state. However, below the $T_0$ the thermal fluctuations are absent, the system is quantum and can be described as 3D one. Then it should come to the ordered state. Perhaps it occurs as an annihilation of the bonded vortices, or the scenario of the discontinuous phase transition characteristic for high-density vortex gas\,\cite{Minnhagen,Ryzhov} is realized. This subject needs additional investigation.

\section*{Appendices}

\section*{Appendix I}

The Hamiltonian has the following form:
\begin{gather*}
H=\sum\limits_{\bf r,l}\mathcal{E}\left[1-\cos( \delta\Phi_{\bf r,\,l})\right],
\end{gather*}
where ${\bf r}$ is a lattice point, ${\bf l}$ are neighboring lattice points, $\mathcal{E}$ is the elastic (exchange) interaction energy.

In the continuous limit making use of the Villain approximation, we can write $H=\int\mathrm{d}^2{\bf r}\mathcal{H}$,
\begin{gather*}
\mathcal{H}=\mathcal{E}(\nabla\Phi)^2.
\end{gather*}

Let us consider the presence in the system of a topological vortex in the point ${\bf r}'$. It means that at a closed loop bypass around this topological vortex the $\nabla\Phi$ gains nonzero value:   \\ $\oint \nabla \Phi \mathrm{d}{\bf l}=\int \nabla\times\nabla\Phi\cdot{\bf J}\mathrm{d}^2{\bf r}=\delta\Phi=\alpha\delta({\bf r-r'}) J\mathrm{d}^2{\bf r}$. Here ${\bf l}$ is the unit of the contour, ${\bf J}$ is the normal to the loop, which contains only $z$-component $J_z=J=\pm 1$, and $\delta\Phi=2\pi$ is the phase shift. In addition, the Hamiltonian contains the vortex core energy $\mathcal{E}_c$.
Therefore the path integral of the system is written as follows:
\begin{gather*}
W=\iint \mathcal{D}\Phi\mathcal{D}{\bf A}
\exp\left[-\hbar^{-1}\tau\int\mathrm{d}^2{\bf r}\,\mathcal{H}\right],
\end{gather*}
where the effective Hamiltonian density of the system with $N$ topological vortices has the following form:
\begin{gather*}
\mathcal{H}=\mathcal{E}(\nabla\Phi)^2+
i\hbar\tau^{-1}{\bf A}[\nabla\times(\nabla\Phi)-2\pi\sum\limits_{n=1}^{N}{\bf J}\delta_{{\bf r=r}_n}]
+\mathcal{E}_c\sum\limits_{n=1}^{N}{\bf J}^2\delta_{{\bf r=r}_n},
\end{gather*}
where ${\bf A}$ is an ancillary field, $\hbar$  is the reduced Planck constant, and $\tau$ is the characteristic time that will be defined below.
After functional integration of
over $\Phi$ we get
\begin{gather*}
\mathcal{H}=\dfr{(\hbar\tau^{-1})^{2}}{4\mathcal{E}}(\nabla\times {\bf A})^2-i\hbar\tau^{-1}2\pi \sum\limits_{n=1}^{N}{\bf A}{\bf J}\delta_{{\bf r=r}_n}+\mathcal{E}_c\sum\limits_{n=1}^{N}{\bf J}^2\delta_{{\bf r=r}_n}.
\end{gather*}
Making use the identity $\nabla\times(\nabla\times{\bf A})\equiv\nabla(\nabla{\bf A})-\nabla^2{\bf A}$, and taking into account the ${\bf A}$ field contains only $z$-component, $A_z=A$, one rewrites recent expression as follows:
\begin{gather}\label{Exp1}
\mathcal{H}=\dfr{(\hbar\tau^{-1})^{2}}{4\mathcal{E}}(\nabla A)^2-i\hbar\tau^{-1} 2\pi \sum\limits_{n=1}^{N}AJ\delta_{{\bf r=r}_n}+\mathcal{E}_c\sum\limits_{n=1}^{N}J^2\delta_{{\bf r=r}_n}.
\end{gather}

In order to take account of all possible vortices configurations we carry out the averaging over a grand canonical ensemble of the ``particles'' endowed with the two possible dimensionless charges: $J_n=\pm 1$.
Then the path integral is:
\begin{multline*}
W=\int\mathfrak{D}A\left\{\exp\left[-\dfr{\tau}{\hbar}\int \mathrm{d}^2{\bf r}\,\dfr{(\hbar\tau^{-1})^{2}}{4\mathcal{E}}
(\nabla A)^2\right] \times\right.\\\left.
\sum\limits_{N=1}^{\infty}\dfr {(e^{-\mathcal{E}_c\tau/\hbar})^N}{N!}\prod\limits_{n=1}^N \int\mathrm{d}^2{\bf r}_n\sum\limits_{J_n=\pm 1}
\exp\left[ i2\pi J_nA({\bf r}_n)\right]\right\}.
\end{multline*}

After averaging over grand canonical ensemble the system effective Hamiltonian density assumes the form:
\begin{gather}
\mathcal{H}=\dfr{\hbar^2}{4\mathcal{E}\tau^2} (\nabla A)^2-\dfr{\hbar g}{\tau a^2}\cos\left(2\pi A\right),
\label{Z1}
\end{gather}
where  $a$ is the characteristic space scale, $g=e^{-\mathcal{E}_c\tau/\hbar}$ is the vortex areal density, and which is nothing but the Hamiltonian density of the sine-Gordon theory\,\cite{Min}.

\section*{Appendix II}

Let us consider the system fluctuation as the result of its interaction with a thermal bath which usually is represented as a system of independent harmonic oscillators. The average energy of the $n\to\infty$ oscillators system is
\begin{gather*}
E_{eq}=\sum\limits_i^n\left[\dfr{\hbar\omega_i}{2}+\dfr{\hbar\omega_i}{e^{\hbar\omega_i/k_bT}-1}\right]
=\dfr12\sum\limits_i^n\hbar\omega_i\,\mbox{cth}\left(\dfr{\hbar \omega_i}{2k_bT}\right),
\end{gather*}
where $\omega_i$ is the eigenfrequency which is the zero-point frequency of i-th oscillator. We suppose that all frequencies are equiprobable. Therefore
\begin{gather*}
E_{eq}=\tau\int\limits_{\omega_0}^{\omega_M}\mathrm{d}\omega\,\dfr{\hbar\omega}2\,\mbox{cth}\left(\dfr{\hbar \omega}{2k_bT}\right),
\end{gather*}
where $\omega_M$ is the Debye frequency, and $\omega_0$ is the minimal zero-point frequency of system fluctuations.

Therefore, in the assumption that the oscillators frequencies evenly distributed from $\omega_0$ till $\omega_M$, in the integral form the average energy is represented as:
\begin{gather*}
E_{eq}=\tau\int\limits_{\omega_0}^{\omega_M}\mathrm{d}\omega\,\dfr{\hbar\omega}2\,\mbox{cth}\left(\dfr{\hbar \omega}{2k_bT}\right)=
\tau\int\limits_{0}^{\infty}\mathrm{d}\omega\,\dfr{\hbar\omega}2\,\mbox{cth}\left(\dfr{\hbar \omega}{2k_bT}\right)\Pi^{\omega_M}_{\omega_0}(\omega),
\end{gather*}
where $\Pi^{\omega_M}_{\omega_0}(\omega)=\theta(\omega-\omega_0)\theta(\omega_M-\omega)$ is the product of the Heaviside functions.

The characteristic time $\tau$ is the time interval at which the system's energy is well determined. It can be estimated from the Heisenberg uncertainty principle since it is the minimal observation time at which the energy uncertainty is less than the fluctuation of energy. Considering the fluctuation of energy as the energy of a single oscillator one can estimate the characteristic time as
\begin{gather*}
\tau=\hbar/E_0=\dfr{2}{\omega_0}\,\mbox{cth}^{-1}\left(\dfr{\hbar \omega_0}{2k_bT}\right).
\end{gather*}
At high temperatures the characteristic time depends on the temperature $\tau =\hbar/k_bT$, and at $T=0$ it is $\tau=2\omega_0^{-1}$.

\section*{Appendix III}

The renormalization of $\lambda$
\begin{gather*}
\lambda^{(R)}=Z_{\lambda}Z_{A^q}Z_{A^{cl}}Z_k^2\Delta^{-1-2/z}=Z_{\lambda}\Delta,
\end{gather*}
where
\begin{multline*}
Z_{\lambda}=\lambda +\sigma^2\dfr{\partial^2}{\partial {\bf k}^2}\left.\langle A^{cl}A^{q}\rangle\right|_{k\to 0}=
\lambda +\sigma^2\left.\int \mathrm{d}^2{\bf r}\,{\bf r}^2 e^{i{\bf rk}}\langle A^{cl}A^{q}\rangle_{\bf r} \right|_{k\to 0}=\\
\lambda +\sigma^2\left.\int\mathrm{d}^2{\bf r}\,{\bf r}^2\langle A^{cl}A^{q}\rangle_{\bf r} \right|_{\omega\to 0}=\lambda +\sigma^2\left.\int\mathrm{d}^2{\bf r}\,{\bf r}^2\int\limits_{{\bf k}^*}^{\Delta^{1/2}{\bf k}^*}\dfr{\mathrm{d}^2{\bf k}}{(2\pi)^2}\dfr{e^{-i{\bf kr}}}{\lambda{\bf k}^2+i\tau\omega}\right|_{\omega\to 0}=\\
\lambda +\dfr{\sigma^2}{\lambda}\int\mathrm{d}^2{\bf r}\,{\bf r}^2\int\limits_{{\bf k}^*}^{\Delta^{1/2}{\bf k}^*}\dfr{\mathrm{d}^2{\bf k}}{(2\pi)^2}\dfr{e^{-i{\bf kr}}}{{\bf k}^2} =\lambda +\dfr{\sigma^2}{4\pi\lambda}\ln\Delta\cdot\int\mathrm{d}^2{\bf r}\,{\bf r}^2 J_0(k^*r)=\\
\lambda +\dfr{\sigma^2}{2\lambda }\ln\Delta\int\limits_0^{\infty}\mathrm{d}r\,r\,r^2 J_0(k^*r)=
\lambda +\dfr{\sigma^2}{2\lambda }\mathcal{H}_0\{r^2\}_{k^*}\ln\Delta,
\end{multline*}
where $\mathcal{H}_0\{r^2\}_{k^*}$ is Hankel transformation of $r^2$ function in $k^*$ point. Therefore
\begin{multline*}
\lambda^{(R)}=Z_{\lambda}Z_{A^q}
Z_{A^{cl}}Z_k^2\Delta^{-1-2/z}=\Delta\left(\lambda +\dfr{\sigma^2}{2\lambda }\mathcal{H}_0\{r^2\}_{k^*}\ln\Delta\right)\\
\approx \lambda+\left(\lambda +\dfr{\sigma^2}{2\lambda }\mathcal{H}_0\{r^2\}_{k^*}\right)\ln\Delta,
\end{multline*}
and
\begin{gather*}
\dfr{d\ln\lambda}{d\ln\Delta}\approx \dfr{\sigma^2}{2\lambda^2 }\mathcal{H}_0\{r^2\}_{k^*},
\end{gather*}
since $\mathcal{H}_0\{r^2\}_{k^*}$ is divergent in the thermodynamic limit.

Similarly, the renormalization of $\alpha$
\begin{gather*}
\alpha^{(R)}=Z_{\alpha}Z_{A^q}Z_{A^{cl}}Z_{\omega}\Delta^{-1-2/z}=Z_{\alpha}\Delta,
\end{gather*}
where
\begin{multline*}
Z_{\alpha}=\alpha +g^2\dfr{\partial}{i\partial \omega}\left.\langle A^{cl}A^{q}\rangle\right|_{k\to 0}=
\alpha -ig^2\left.\int \mathrm{d}t\,t e^{it\omega}\langle A^{cl}A^{q}\rangle_{t} \right|_{k\to 0}=\\
\alpha -ig^2\left.\int\mathrm{d}t\,t\langle A^{cl}A^{q}\rangle_t \right|_{\bf k\to 0}=\alpha -ig^2\left.\int\mathrm{d}t\,t\int\limits_{\omega^*}^{\Delta\omega^*}
\dfr{\mathrm{d}\omega}{2\pi}\dfr{e^{-it\omega}}{\lambda{\bf k}^2+i\alpha\omega}\right|_{{\bf k}\to 0}=\\
\alpha -\dfr{ig^2}{\alpha}\int\mathrm{d}t\,t\int\limits_{\omega^*}^{\Delta\omega^*}
\dfr{\mathrm{d}\omega}{2\pi}\dfr{e^{-it\omega}}{i\omega} =\alpha -\dfr{g^2}{ 2\pi\alpha}\int\limits_{\omega^*}^{\Delta\omega^*}
\dfr{\mathrm{d}\omega}{\omega}\int\mathrm{d}t\,t e^{-it\omega}=\\
\alpha -\dfr{g^2}{\alpha}\int\limits_{\omega^*}^{\Delta\omega^*}
\mathrm{d}\omega\dfr{i}{\omega}\delta^{(1)}(\omega)=\alpha.
\end{multline*}
Thus
\begin{gather*}
\dfr{d\ln\alpha}{d\ln\Delta}\approx 1.
\end{gather*}

\section*{Appendix IV (The $d=2$ case)}

The key problem in the renormalization of $\sigma$ close to $T=0$ is the calculation of \begin{gather*}
Z_{\sigma}\approx\sigma\exp\left[-\dfr{(\sqrt{2}\pi)^2}2 \Omega(\Delta)\right],
\end{gather*}
where
\begin{gather*}
\Omega(\Delta)\propto
\int\limits_{\omega^*}^{\Delta\omega^*}\mathrm{d}\omega\,\dfr{\omega \,\mbox{cth}\left({\hbar\omega}/{2k_bT}\right)\Pi^{\omega_M}_{\omega_0}(\omega)}{\omega}.
\end{gather*}

\begin{figure}
\centering
   \includegraphics[scale=1.2]{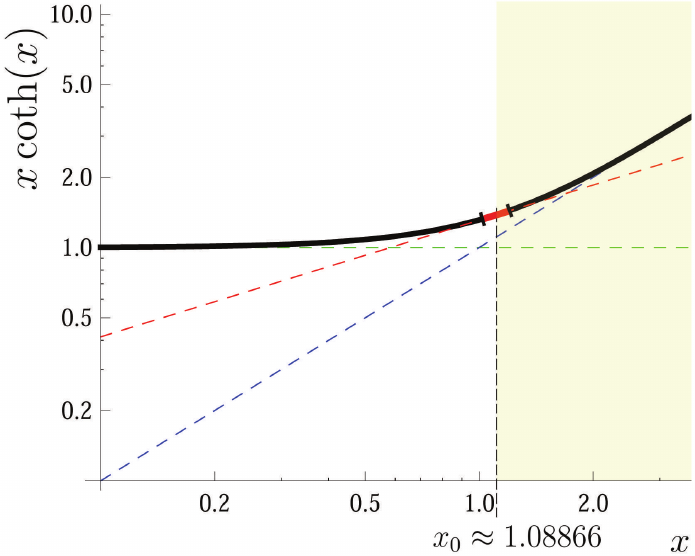}
   \caption{ The thick black line is the log--log plot of the $x\,\mbox{cth}(x)$ function. Red dashed line is the tangent to $x\,\mbox{cth}(x)$ function in $x=x_0$ point, blue dashed line is the tangent to $x\,\mbox{cth}(x)$ function at $x\gg 1$, and green dashed line is the tangent to $x\,\mbox{cth}(x)$ function at $x\to 0$ point. In neighborhood of $x=x_0$ the $x\,\mbox{cth}(x)$ function  (the thick red linear segment) can be approximated by the exponent function $\Lambda \ln x +A$ ($\Lambda\approx 1/2$) whose log--log plot is the tangent line to $x\,\mbox{cth}(x)$.}
   \label{LogPlot}
\end{figure}

In order to reduce the renormalization procedure to the standard form one should approximate function $x\,\mbox{cth}(x)$ by an exponential function, $ x^{\Lambda (x)}$.
Note that critical dynamics considers a system in $\omega \to 0$ limit. However, this theoretical limit is unreachable practically. The natural limits are observation time or scale of zero-point fluctuations time in quantum case.
Let us consider this approximation near to the lower limit of the frequency scale, $\omega = \omega_0$, that is relevant for critical dynamics. In the Fig.\,\ref{LogPlot} the $x\,\mbox{cth}(x)$ function is represented in logarithmic coordinates. The sought-for exponential approximation of this function in some point $x=x^*$ is the tangent to this point. It is $\ln (x\,\mbox{cth}(x))=\Lambda (x^*)\ln x +A(x^*)$, where $\Lambda (x^*)=\left.\partial \ln(x\,\mbox{cth}(x))/\partial \ln x \right|_{x=x^*}$, and $A(x^*)=\ln\left({x^*}^{1-\Lambda (x^*)}\mbox{cth}(x^*)\right)$. Therefore at the point $x=x_0$ the approximation has the following form:
\begin{gather}
\left. x\,\mbox{cth}(x)\right|_{x=x^*}\propto x^{\Lambda (x^*)}\exp A(x^*),
\end{gather}
where
\begin{gather*}
\Lambda (x^*) = 1-2x^*\mbox{csch}(2x^*).
\end{gather*}
This is single exponential approximation the considered function in given point.

Taking into account the above reasoning $\Omega(\Delta)$ can be calculated as follows:
\begin{multline*}
\Omega(\Delta)=\dfr2{(2\pi)^3}\int\limits_{k^*}^{\Delta k^*}\mathrm{d}^3k\langle A^{cl}A^{cl}\rangle_{k}=
\int\limits_{-\infty}^{\infty}\dfr{\mathrm{d}^2{\bf k}}{(2\pi)^2}\int\limits_{\omega_0}^{\Delta\omega_0}\dfr{\mathrm{d}\omega}{2\pi}\,\dfr{2\alpha\omega \,\mbox{cth}\left({\hbar\omega}/{2k_bT}\right)}{\lambda^{2}{\bf k}^4+\alpha^2\omega^2}=\\
\dfr{\alpha}{\lambda\alpha}\int\limits_{-\infty}^{\infty}\dfr{\mathrm{d}^2{\bf p}}{(2\pi)^2}\int\limits_{\omega_0}^{\Delta\omega_0}\dfr{\mathrm{d}\omega}{2\pi}\,\dfr{2\omega \,\mbox{cth}\left({\hbar\omega}/{2k_bT}\right)}{{\bf p}^4+\omega^2}=
\dfr{1}{4\lambda}\int\limits_{\omega_0}^{\Delta\omega_0}\dfr{\mathrm{d}\omega}{2\pi}\,\dfr{\omega \,\mbox{cth}\left({\hbar\omega}/{2k_bT}\right)}{\omega}=\\
\dfr{{k_bT}}{4\pi\hbar\lambda}\int\limits_{\omega_0{\hbar}/{2 k_bT}}^{\Delta\omega_0{\hbar}/{2 k_bT}}\mathrm{d}x\,\dfr{x \,\mbox{cth}\left(x\right)}{x}\approx
\dfr{{k_bT}}{4\pi\hbar\lambda}{x_0}^{1-\Lambda(x_0)}\,\mbox{cth}\left(x_0\right)
\int\limits_{x_0}^{\Delta x_0}\mathrm{d}x\,x^{\Lambda(x_0)-1}=\\
\dfr{\omega_0}{8\pi\lambda}\,\dfr{\mbox{cth}\left(x_0\right)}{{x_0}^{\Lambda(x_0)}}
\int\limits_{ x_0}^{\Delta x_0}\mathrm{d}\ln x\,x^{\Lambda(x_0)}=
\dfr{\omega_0}{8\pi\lambda}\,\dfr{\mbox{cth}\left(x_0\right)}{{x_0}^{\Lambda(x_0)}}
\int\limits_{\ln x_0}^{\ln x_0+\ln\Delta}\mathrm{d}y\,e^{\Lambda(x_0)y}=\\
\dfr{\omega_0}{8\pi\lambda}\,\dfr{\mbox{cth}\left(x_0\right)}{\Lambda(x_0){x_0}^{\Lambda(x_0)}}
e^{\Lambda(x_0)\ln x_0}\left(e^{\Lambda(x_0)\ln\Delta}-1\right)=\\
\dfr{\omega_0}{8\pi\lambda}\dfr{\mbox{cth}\left(x_0\right)}{\Lambda(x_0){x_0}^{\Lambda(x_0)}}
{x_0}^{\Lambda(x_0)}\left(\Delta^{\Lambda(x_0)}-1\right)=\dfr{\omega_0}{8\pi\lambda}
\dfr{\mbox{cth}\left(x_0\right)}{\Lambda(x_0)}\left(e^{\Lambda(x_0)\ln\Delta}-1\right),
\end{multline*}
where $x_0=\omega_0{\hbar}/{2 k_bT}$, and $\Lambda (x_0) = 1-2x_0\mbox{csch}(2x_0)$. Also one can estimate the temperature of the crossover between thermal and quantum critical dynamics regimes: $T_0={\hbar\omega_0}/{2k_bx_0}$.

\end{document}